\documentclass[preprint]{aastex}

\usepackage{natbib}
\usepackage{txfonts}
\usepackage{subfigure}

\bibpunct{(}{)}{;}{a}{}{,}

\shorttitle{A method to identify and characterise binary candidates}
\shortauthors{R. Da Silva \& A. Silva-Valio}

\begin{document}

\def\msun{${\rm M}_{\odot}$} 
\def\rsun{${\rm R}_{\odot}$} 
\def\mjup{${\rm M}_{\rm Jup}$}
\def\rjup{${\rm R}_{\rm Jup}$}
\def\m{$-$}

\title{A method to identify and characterise binary candidates - \\
       a study of CoRoT\thanks{\,The CoRoT space mission, launched on 2006
       December 27, was developed and is operated by the CNES, with
       participation of the Science Programs of ESA, ESA's RSSD, Austria,
       Belgium, Brazil, Germany and Spain.}\,\,\,data}

\author{Ronaldo Da Silva}
\affil{Divis\~ao de Astrof\'\i sica, Instituto Nacional de Pesquisas
Espaciais, S\~ao Jos\'e dos Campos, Brazil}
\email{dasilvr2@gmail.com}
\and
\author{Adriana Silva-Valio}
\affil{Centro de R\'adio-Astronomia e Astrof\'\i sica Mackenzie,
Universidade Presbiteriana Mackenzie, S\~ao Paulo, Brazil}

%
%
\begin{abstract}
The analysis of the CoRoT space mission data was performed aiming to test a
method that selects, among the several light curves observed, the transiting
systems that likely host a low-mass star orbiting the main target. The
method identifies stellar companions by fitting a model to the observed
transits. Applying this model, that uses equations like Kepler's third law
and an empirical mass-radius relation, it is possible to estimate the mass
and radius of the primary and secondary objects as well as the semimajor
axis and inclination angle of the orbit. We focus on how the method can be
used in the characterisation of transiting systems having a low-mass stellar
companion with no need to be monitored with radial-velocity measurements or
ground-based photometric observations. The model, which provides a good
estimate of the system parameters, is also useful as a complementary
approach to select possible planetary candidates. A list of confirmed
binaries together with our estimate of their parameters are presented. The
characterisation of the first twelve detected CoRoT exoplanetary systems was
also performed and agrees very well with the results of their respective
announcement papers. The comparison with confirmed systems validates our
method, specially when the radius of the secondary companion is smaller than
1.5~\rjup, in the case of planets, or larger than 2~\rjup, in the case of
low-mass stars. Intermediate situations are not conclusive.
\end{abstract}


%
%
\section{Introduction}

The analysis of transiting systems based on light curve modelling combined
with ground-based follow-up by means of radial-velocity measurements have
shown their exceptional importance in the characterisation of extrasolar
systems. Together, they provide the determination of several physical and
orbital parameters that cannot normally be obtained for non-transiting
systems, such as the mass and radius of the secondary companion, and thus
its density. The possibility of a detailed study of extrasolar transiting
light curves in a large number of targets has been considerably improved by
recent photometric space missions, such as CoRoT \citep{Baglinetal2006} and
Kepler \citep{Boruckietal2010}.

At the time of this writing the CoRoT space telescope has collected, since
its launch, the light curves of more than a hundred thousand stars through
13 observational runs. This huge number of stars is a strong motivation to
develop tools to efficiently treat the released data, specially considering
that first of all the data need to be cleaned from long-term variations,
short-term oscillations, outliers, discontinuities, and others. Regarding
the characterisation of extrasolar systems, an extra challenge is to
distinguish transits caused by planetary companions from those related to
the presence of a low-mass star in a binary system, particularly given the
time-consuming ground-based observations that are normally used as a
complementary approach.

\citet{Carpanoetal2009} published their results concerning the analysis of
CoRoT light curves observed during the initial run of the mission, named
IRa01\footnote{\,IR: {\it Initial Run}; LR: {\it Long Run}; SR: {\it Short
Run}; {\it c} and {\it a} represent the direction of the galactic and the
anti-galactic centre, respectively.}, in which they presented a list of 50
planetary candidates together with a list of 145 eclipsing binary
candidates. \citet{Moutouetal2009} complemented the work of
\citet{Carpanoetal2009} with additional follow-up observations, also in the
context of the initial run. In other two papers, published by
\citet{Deegetal2009} and \citet{Cabreraetal2009}, the authors presented a
list of targets for which ground-based follow-up was conducted, helping as a
complementary approach in the classification of the candidates. The lists
released by these papers are the result of a huge effort of several working
groups, and shows how difficult it is to characterise systems, planetary
candidates or not, among thousands of targets.

In the work of \citet{SilvaCruz2006}, the authors proposed a method based on
the fit of light curves with transits that can be used in the
characterisation of transiting systems. This method provides the
determination of some parameters of the system, such as the orbital
inclination angle, the semimajor axis, and the mass and radius of the
primary and secondary objects. In the present work we have used an updated
version of the same method, applied to a list of transiting light curves
from publicly available runs of the CoRoT mission. The purpose of our
analysis is to show that the method is useful to identify transits most
likely caused by a binary configuration, without making use of any
time-consuming effort to conduct ground-based follow-up, like radial
velocity measurements or photometric observations.

\citet{SilvaCruz2006} also tested their model applied to the transiting
exoplanetary systems known at that time, and their results are in good
agreement with the published parameters. Therefore, here we analysed the
first confirmed CoRoT planetary systems as well, from CoRoT-1 through 12,
comparing our estimates to those in their respective announcement papers.

Section~\ref{reduction} presents the data reduction, showing the corrections
needed to apply to the light curves (in the format delivered to the
scientific community) before modelling the transit shape. Next,
Sect.~\ref{method} describes the method and its usefulness in the
classification of binary system candidates. In Sect.~\ref{results},
the results are presented and discussed, which includes our parameter
estimate for a list of binary systems first identified by other works as
well as the characterisation of confirmed CoRoT exoplanetary systems.
Finally, Sect.~\ref{concl} presents the final remarks and conclusions.

%
%
\section{Sample data and reduction}
\label{reduction}

The light curves analysed using our method are part of the three-colour band
data from publicly available runs observed by the CoRoT mission (a few cases
of monochromatic light curves were also included). Before doing any kind of
fit to model the observed transits, the light curves have to be cleaned from
any systematic noise that may still remain in the data delivered to the
scientific community, which format is called the N2 level
\citep{Baudinetal2006}. The systematic noise normally seen is related to:
$i)$ discontinuities produced by hot pixels;
$ii)$ outliers, whose sources are diverse; and
$iii)$ short-term oscillations related to the CoRoT orbital frequency
(103 min) and its harmonics. For details on the CoRoT satellite and its
orbit see e.g. \citet{BoisnardAuvergne2006} and \citet{Auvergneetal2009}.
In addition to the systematic noise, short-term oscillations, intrinsic to
some type of stars, were also observed in some light curves and removed. 

\begin{figure}[t!]
\centering
\resizebox{0.4\hsize}{!}{\includegraphics{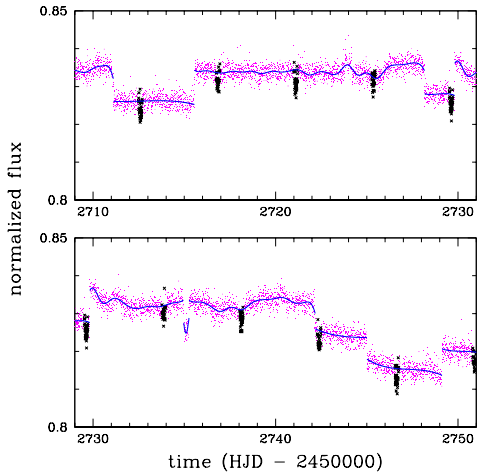}}
\caption{Example of polynomial function fit performed to correct the light
         curves from eventual discontinuities. Only points within twice
	 $\sigma$ and located outside transits (dots) are taken into account
	 when searching for the best fit (solid lines). Transit positions
	 are represented by crosses. This figure represents CoRoT-1, which
	 folded light curve is shown in Fig.~\ref{corot_fig} after the
	 discontinuity corrections were made.}
\label{discont}
\end{figure}

We developed a code to correct the light curves from discontinuities and
outliers. The code computes the first derivate of the data points in order
to locate the discontinuities, then it fits polynomial functions to points
between them, and finally the light curve is normalised according to such
functions. When searching for the best fit, the code does not consider:
$i)$ data points located outside twice $\sigma$, which value is estimated
in a specific portion of the light curve; and
$ii)$ data points located where transits happen.
We note that a visual inspection is always conducted to avoid any abrupt
behaviour of the fitted function, specially close to transit times.
Figure~\ref{discont} shows an example of this polynomial function fit.

\begin{figure}[t!]
\centering
\resizebox{0.4\hsize}{!}{\includegraphics{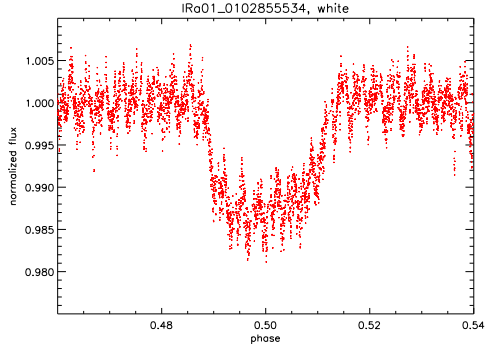}}
\caption{Phase-folded light curve with short-term oscillations (in this case
        due to the CoRoT orbit) that we have to correct before modelling the
	transit shape. The same system is shown in Fig.~\ref{bin_cand} after
	these oscillations have been removed.}
\label{oscill}
\end{figure}

Short-term oscillations may also affect the search for the model that best
fits the observed transit. Figure~\ref{oscill} shows an example of a
transiting light curve when this kind of oscillation is present. The same
object is shown in Fig.~\ref{bin_cand} after being corrected from the most
prominent harmonics. Oscillations due to the intrinsic variability of some
stars were also properly removed when needed.

%
%
\section{Light curve fit}
\label{method}
The method developed by \citet{SilvaCruz2006} was used here to search for
the best model parameters that fit the observed light curve transits. The
model considers an opaque disc that simulates the secondary object passing
across the stellar disc. We assumed a quadratic function to describe the
limb-darkening of the disc of the primary object, which is based on the star
HD\,209458 \citep[$u_1 = 0.2925$ and $u_2 = 0.3475$, from][]{Brownetal2001}.
An exception is the system 0100773735 (LRc01, Fig.~\ref{bin_cand}), for
which $u_1$ and $u_2$ were assumed to be half those of the star HD\,209458.
In the case in which the secondary companion is a low-mass star, it will not
be an opaque disc. However, this will not considerably change the results,
since its flux contribution is small compared to the main star (a 0.3~\msun\
star orbiting a solar-type star contributes less than 2\% to the total
flux). Cases in which the secondary is as bright as the primary were not
selected in our analysis given the large transit depth that would be
observed in the light curve.

The orbital period ($P$) of the companion is a known parameter, obtained
directly from the light curve, whereas the following three variables are the
result of the best fit: the radii ratio between secondary and primary
objects \mbox{($R_{\rm p} = R_2/R_1$)}, the semimajor axis of the secondary
orbit in units of the primary radius ($a_{\rm p} = a/R_1$), and the orbital
inclination angle ($i$). The search for the best fit is conducted with the
AMOEBA routine \citep{Pressetal1992}, which performs a multidimensional
chi-square ($\chi^2$) minimisation of the function $f$($R_{\rm p}$,
$a_{\rm p}$, $i$) describing the transit profile.

\begin{figure*}[t!]
\centering
\resizebox{0.7\hsize}{!}{\includegraphics[angle=-90]{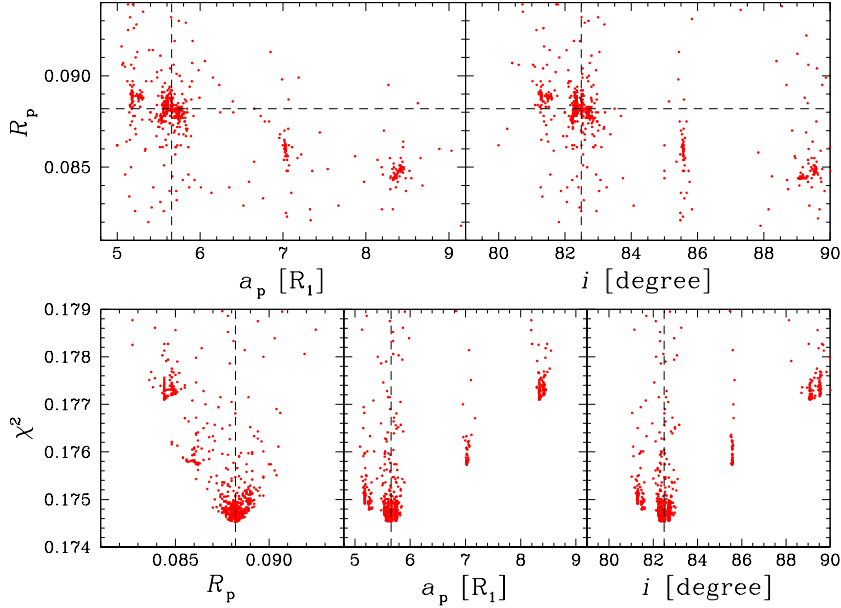}}
\caption{Example of $\chi^2$ minimisation for the system 0100773735 (LRc01,
         Fig.~\ref{bin_cand}) in the search for values of $R_{\rm p}$,
	 $a_{\rm p}$, and $i$ of the model that best fit the observed
	 transit.}
\label{chi2}
\end{figure*}

The first set of parameters normally used as input are: $R_{\rm p}$ = 0.1,
$i = 85^\circ$, and $a_{\rm p} = a/R_1$ calculated according to the Kepler's
third law (Equation~\ref{kepler}) for $M_1 + M_2 \sim 1$~\msun\ and
$R_1 \sim 1$~\rsun. In order to avoid premature convergences on local
minimums and explore the entire parameter space, the best solution is found
after running the routine several times, replacing the input parameters by
new values within a chosen range (e.g. $\Delta R_{\rm p} = \pm 0.05$,
$\Delta a_{\rm p} = \pm 3.0~R_1$, and $\Delta i = \pm 5^\circ$). Throughout
the execution of the process, a visual inspection of the fit is carried out
after each possible solution is achieved. Figure~\ref{chi2} shows an example
of $\chi^2$ minimisation for the system 0100773735 (LRc01).


\subsection{Estimate of mass and radius}
\label{mass_radius}

\begin{figure}[t!]
\centering
\resizebox{0.4\hsize}{!}{\includegraphics{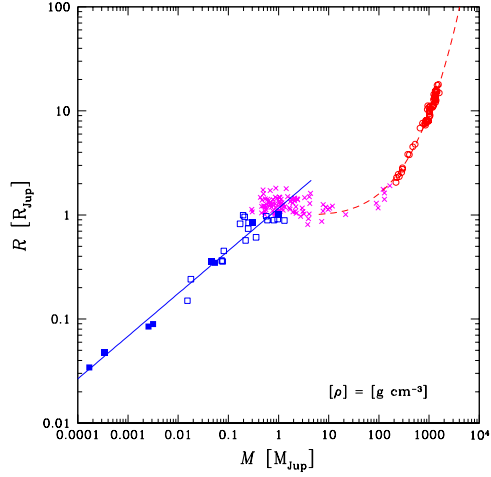}}
\caption{Empirical mass-radius relations used by our method. The long-dashed
         line is a function, given by Equation~\ref{eq_mr_1}, fitted to
	 stars (open circles) whereas the solid line represents a function,
	 given by Equation~\ref{eq_mr_2}, fitted to known exoplanets (open
	 squares) and the planets of the Solar System (filled squares).
	 Crosses represent exoplanets and low-mass objects that were not
	 used in the fit (see text). The values of mass and radius were
	 taken from the Extrasolar Planets Encyclopaedia$^3$.}
\label{massradius}
\end{figure}

The method used to estimate the mass and radius of the primary and secondary
companion was updated using new mass-radius relations based on more recent
discoveries, specially the one fitted to known exoplanets (systems presently
listed in the
Extrasolar Planets Encyclopaedia\footnote{\,http://exoplanet.eu} were used).
These relations are:
\begin{equation}
\label{eq_mr_1}
R = A{M}^B
\end{equation}
\begin{equation}
\label{eq_mr_2}
R = C{M} + D
\end{equation}
where $A = 0.013938$, $B = 3.867270$, $C = 0.410162$, and $D = 0.065934$.
The function given by Equation~\ref{eq_mr_1} was fitted to stars whereas
Equation~\ref{eq_mr_2} represents a function fitted to known exoplanets and
the planets of the solar system. Both functions are plotted in
Fig.~\ref{massradius}. Exoplanets having radius $> 1$~\rjup\ or mass
$> 2$~\mjup\ and low-mass objects having radius $< 2$~\rjup\ were not used
in the search for the best fit. These objects belong to an ambiguous region
of the mass-radius diagram, in which a change in mass does not necessarily
imply in a change in radius.

After applying our method to the observed transits, we are able to estimate
the mass and radius of both the primary ($M_1$, $R_1$) and secondary
($M_2$, $R_2$) objects and the orbital semimajor axis $a$ (given in
astronomical units) using the following relations:
\begin{equation}
\label{kepler}
a^3 = \frac{GP^2}{4\pi^2}(M_1 + M_2)
\end{equation}
\begin{equation}
\label{eq_Rp}
R_{\rm p} = \frac{R_2}{R_1}
\end{equation}
\begin{equation}
\label{eq_ap}
a_{\rm p} = \frac{a}{R_1}
\end{equation}
\begin{eqnarray}
\label{sol_1}
R_1  =  A{M_1}^B,  & &  R_2 = A{M_2}^B
\end{eqnarray}
\begin{eqnarray}
\label{sol_2}
R_1  =  A{M_1}^B,  & &  R_2 = C{M_2} + D
\end{eqnarray}
where Equation~\ref{kepler} is the Kepler's third law, Equations~\ref{eq_Rp}
and \ref{eq_ap} are obtained by the transit best fit, and
Equations~\ref{sol_1} and \ref{sol_2} are the empirical mass-radius
relations used depending on whether the secondary companion is a star or a
planet. Therefore, two sets of five relations are numerically solved and two
sets of the parameters $M_1$, $R_1$, $M_2$, $R_2$, and $a$ are computed.
If $R_2 \gtrsim 2$~\rjup\ then we consider the system to be a binary
candidate and the parameters yielded by Equations~\ref{kepler}, \ref{eq_Rp},
\ref{eq_ap}, and \ref{sol_1} are used. On the other hand,
if $R_2 < 2$~\rjup, then the parameters yielded by Equations~\ref{kepler},
\ref{eq_Rp}, \ref{eq_ap}, and \ref{sol_2} are used.

The numerical calculation of $M_1$, $R_1$, $M_2$, $R_2$, and $a$ proceeds as
follows: first, one of the parameters is fixed (e.g. $M_1$ = 1~\msun) and
the others are calculated; then, the ratio $a/R_1$ is compared to the value
of $a_{\rm p}$ provided by the transit best fit; the fixed parameter is
iteratively incremented (or decremented) until the difference between
$a/R_1$ and $a_{\rm p}$ is as smaller as one wishes. When this condition is
satisfied, the five parameters are finally estimated.

Concerning the mass of the secondary objects ($M_2$), there are three
possibilities:
$i)$ if the companion has $R_2 > 2$~\rjup, then it will probably be a
low-mass star and its mass can be easily estimated by the mass-radius
relation fitted to that region of the diagram;
$ii)$ if $R_2 < 1$~\rjup, then we can estimate a value for $M_2$ using the
other mass-radius relation; however, since the mass-radius relation for
planets depends on their chemical composition, it is not possible to
estimate an accurate mass value for the secondary in this radius regime if
its composition is unknown;
$iii)$ finally, if $1 \leq\ R_2 \leq\ 2$~\rjup, then the mass can not be
univocally determined since, for one given radius, objects in this radius
regime may have masses ranging from about $1~$\mjup\ to about $100~$\mjup\
(brown dwarfs).
Based on these arguments, we only include in our discussion a mass estimate
for secondary companions with $R_2 \gtrsim 2 $~\rjup. Nevertheless, we
note that our method provides a good estimate for the radius of not only
primary and secondary stars, but also of planetary companions (even for
$R_2 < 2$~\rjup). Indeed, this is the case of the confirmed CoRoT systems,
which results are presented and discussed in Sect.~\ref{results}.


\subsection{Uncertainties in the system parameters}
\label{errors}

The results presented here are for the parameters $a$, $i$, $M_1$, $R_1$,
$M_2$, and $R_2$, where $a$ is given in astronomical units, $M_1$ and $M_2$
in solar masses, $R_1$ in solar radii, and $R_2$ in Jupiter radii. To
estimate the uncertainty in these parameters, the procedure is as follows:

\begin{itemize}

\item[\it 1)] first, the standard deviation ($\sigma$) is computed in a
region outside a transit for the model that best fits the light curve; in a
region inside a transit, a possible presence of spots on the surface of the
primary star would cause variations in the light curve \citep{Silva2003} and
$\sigma$ would be miscalculated;

\item[\it 2)] by changing, at a given step, the three basic parameters
yielded from the best fit to the transit ($R_{\rm p}$, $a_{\rm p}$, and
$i$), we obtain a new light curve;

\item[\it 3)] next, these parameters are changed until the new light curve
starts deviating significantly from the best one; that is when the
difference between the light curves is more than 1~$\sigma$ for at least
10\% of the transit;

\item[\it 4)] finally, these new basic parameters are used to re-estimate
$M_1$, $M_2$, $R_1$, and $R_2$ following the procedure described in
Sect.~\ref{mass_radius}; the difference between both new and best estimates
provides the uncertainties in the parameters.

\end{itemize}

Our mass-radius relations (Equations~\ref{eq_mr_1} and \ref{eq_mr_2}) do not
depend directly on the orbital inclination angle. However, since a change in
this parameter leads to a change in both depth and duration of the transit,
its uncertainty can be included in the uncertainty determination of the
other two basic parameters. To do so, first the uncertainty in the
inclination angle is estimated as described above (in steps 2 and 3 only
$i$ is changed). Then, the uncertainties in the other two parameters (each
one in turn) are estimated considering the inclination angle changed by its
uncertainty. At the end, the uncertainties in $M_1$, $R_1$, $M_2$, and $R_2$
take into account those in $R_{\rm p}$, $a_{\rm p}$, and $i$. The
uncertainty in $a$ includes those in both $a_{\rm p}$ and $R_1$ given the
conversion of the semimajor axis from units of stellar radius to
astronomical units.

To reduce noise and clarify visualisation, a smoothing function was applied
to the light curves before searching for the model that best fits the
transit. In addition, for some systems only the best quality channels were
used in the analysis. These procedures result in a more accurate estimate of
the system parameters and, consequently, in smaller uncertainties.

%
%
\section{Results and discussion}
\label{results}

\subsection{Characterisation of binary system candidates}
\label{res_bin}

\begin{figure*}[p!]
\centering
\begin{minipage}[t]{0.49\textwidth}
\centering
\resizebox{\hsize}{!}{\includegraphics{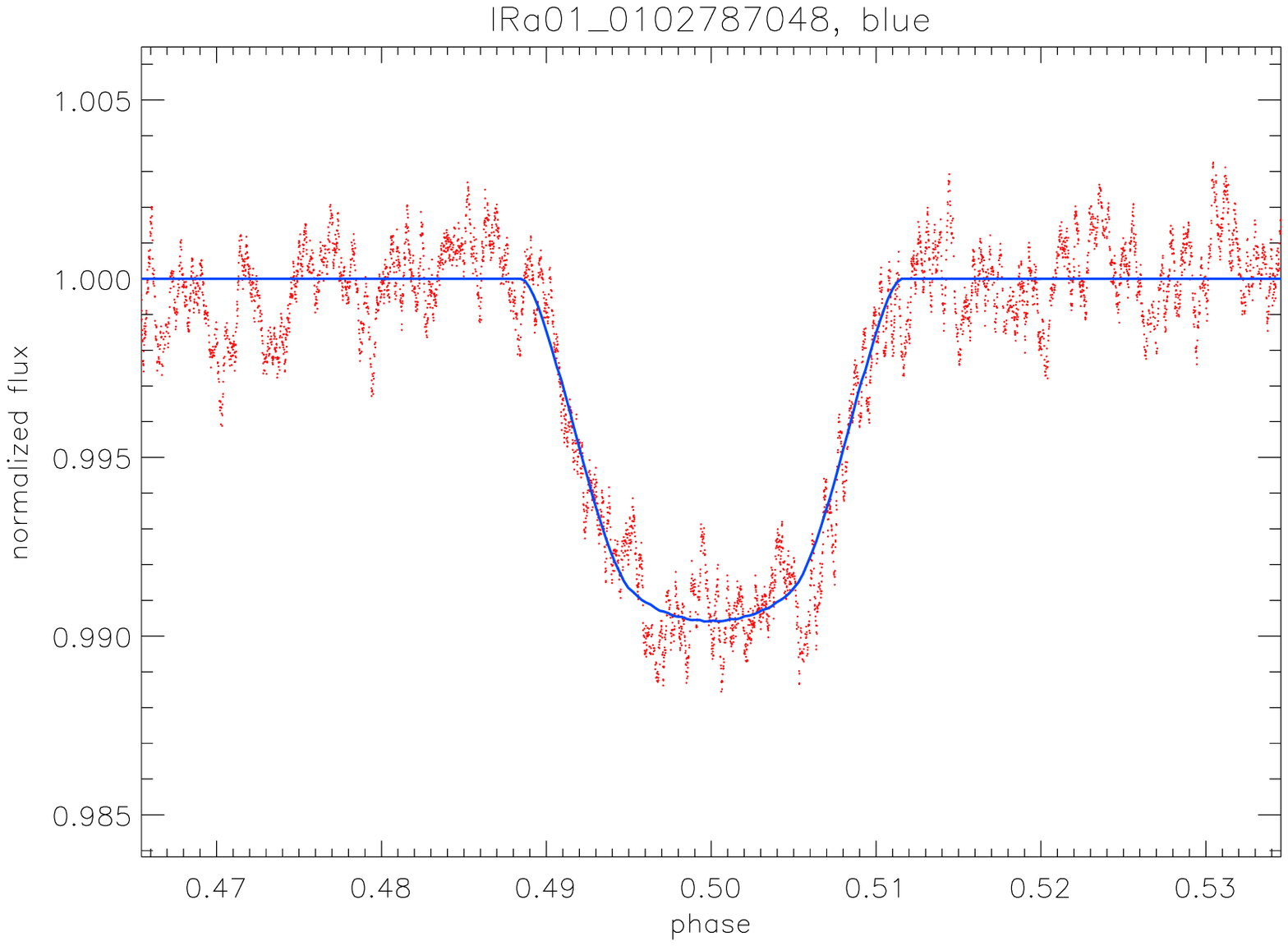}}
\end{minipage}
\begin{minipage}[t]{0.49\textwidth}
\centering
\resizebox{\hsize}{!}{\includegraphics{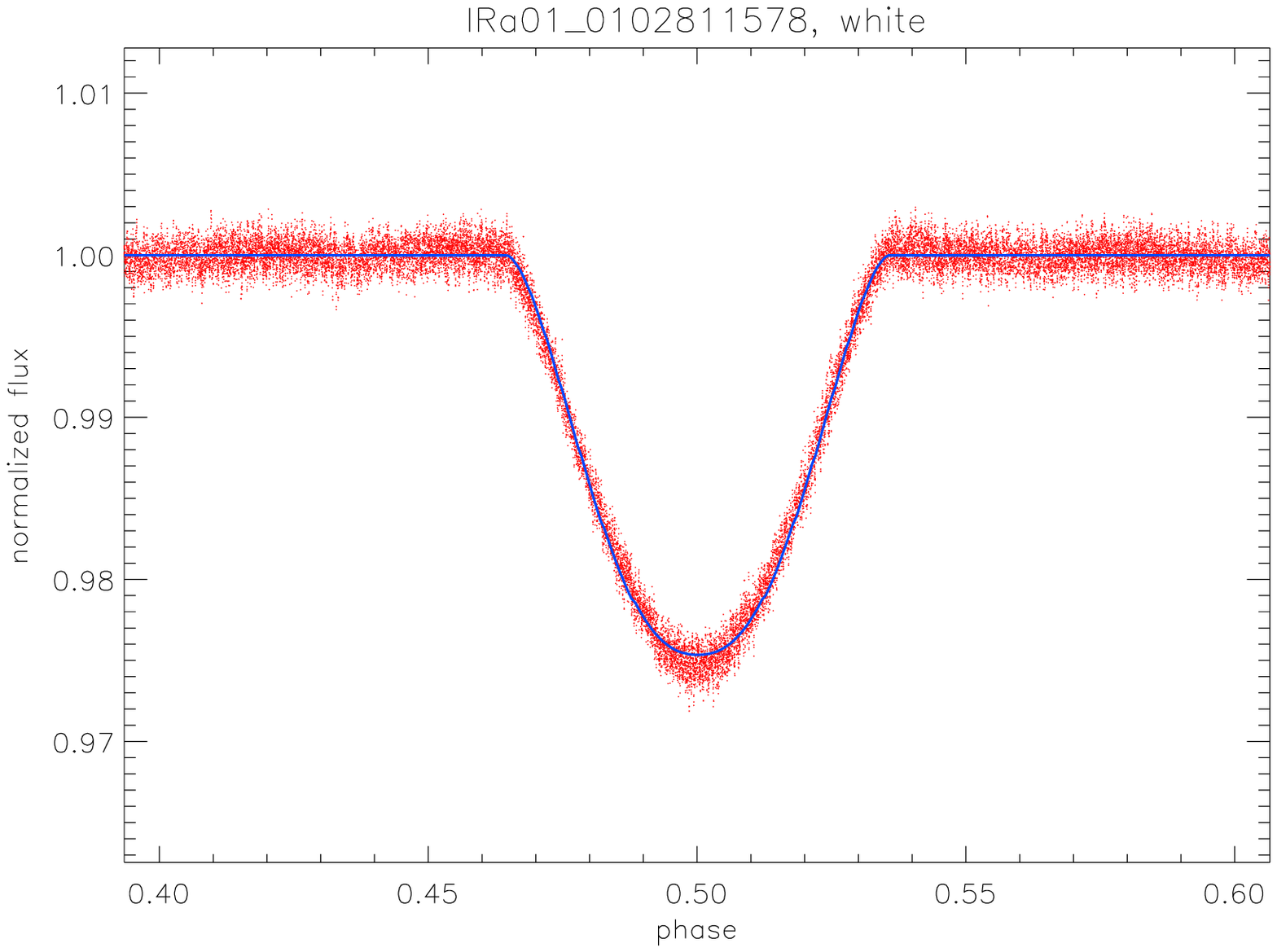}}
\end{minipage} \\[0.80cm]
\begin{minipage}[t]{0.49\textwidth}
\centering
\resizebox{\hsize}{!}{\includegraphics{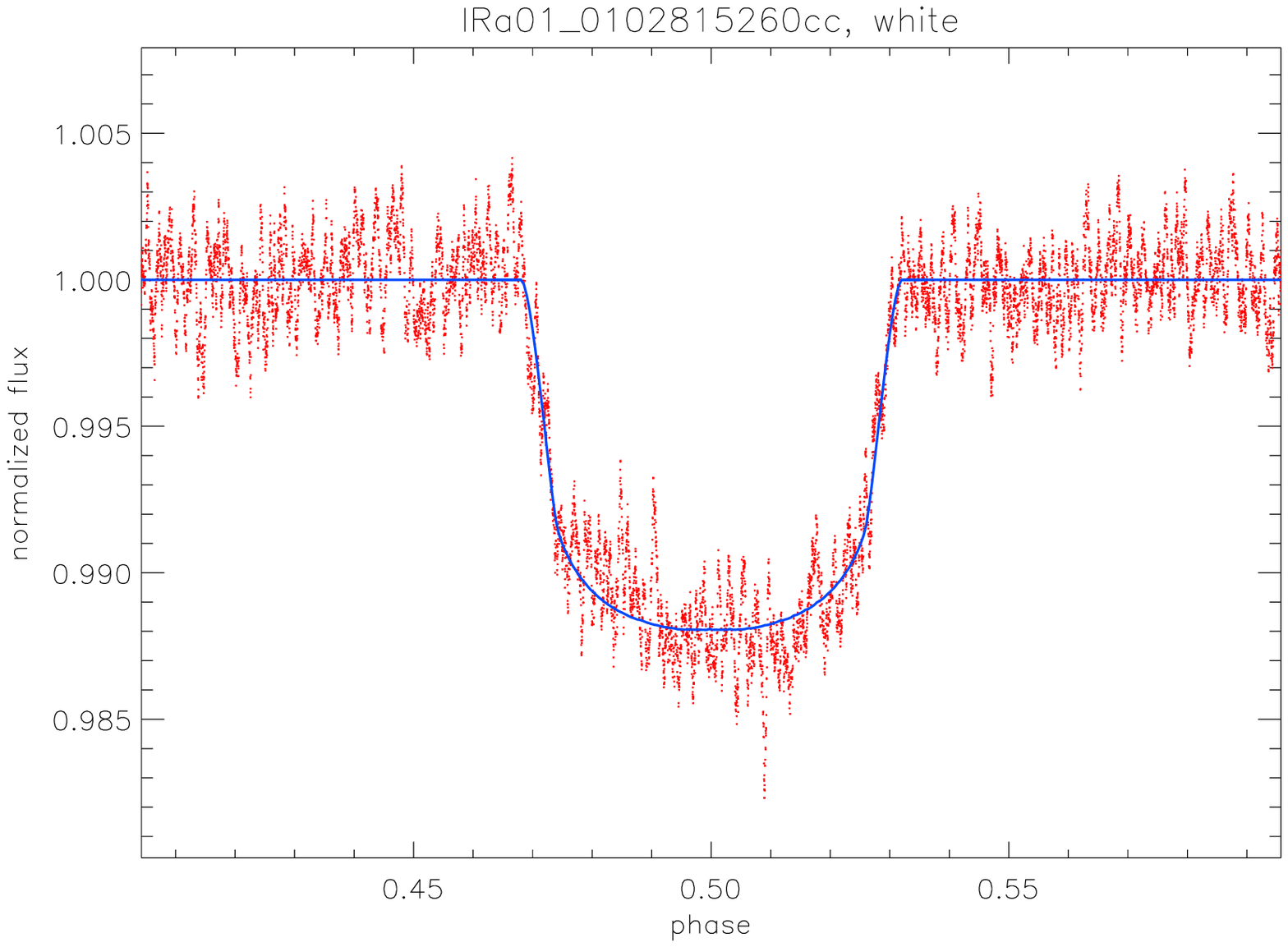}}
\end{minipage}
\begin{minipage}[t]{0.49\textwidth}
\centering
\resizebox{\hsize}{!}{\includegraphics{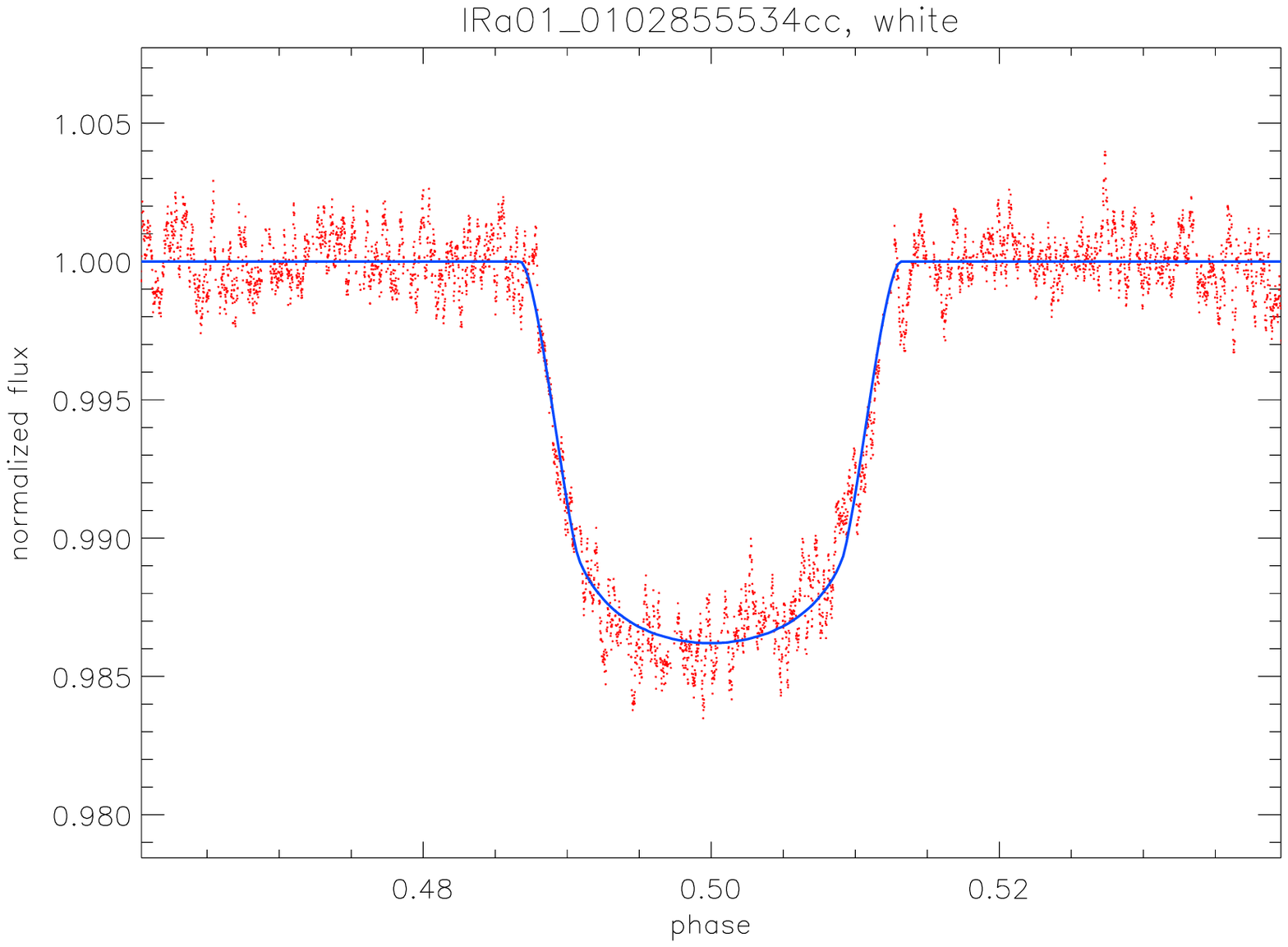}}
\end{minipage} \\[0.80cm]
\begin{minipage}[t]{0.49\textwidth}
\centering
\resizebox{\hsize}{!}{\includegraphics{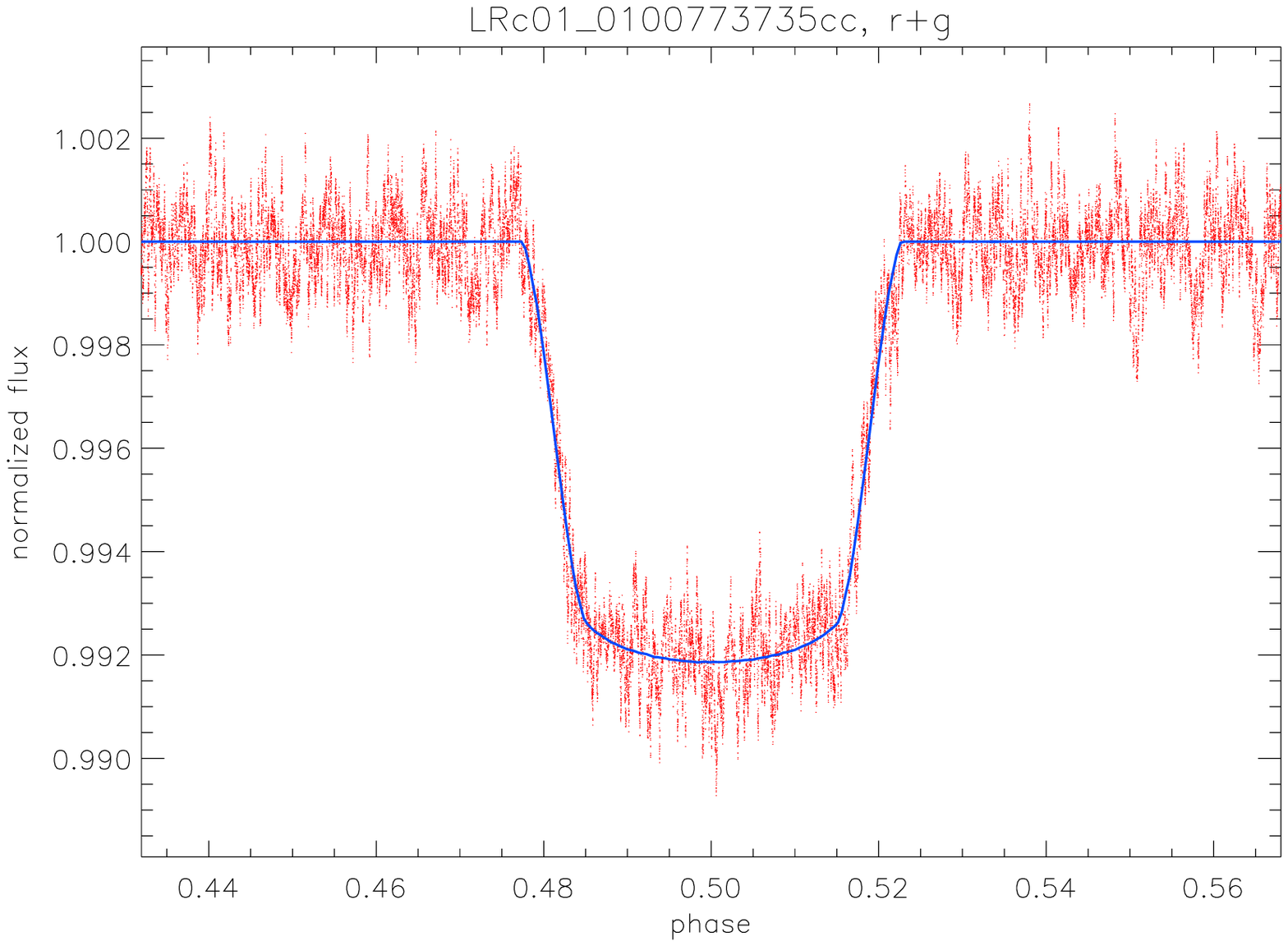}}
\end{minipage}
\begin{minipage}[t]{0.49\textwidth}
\centering
\resizebox{\hsize}{!}{\includegraphics{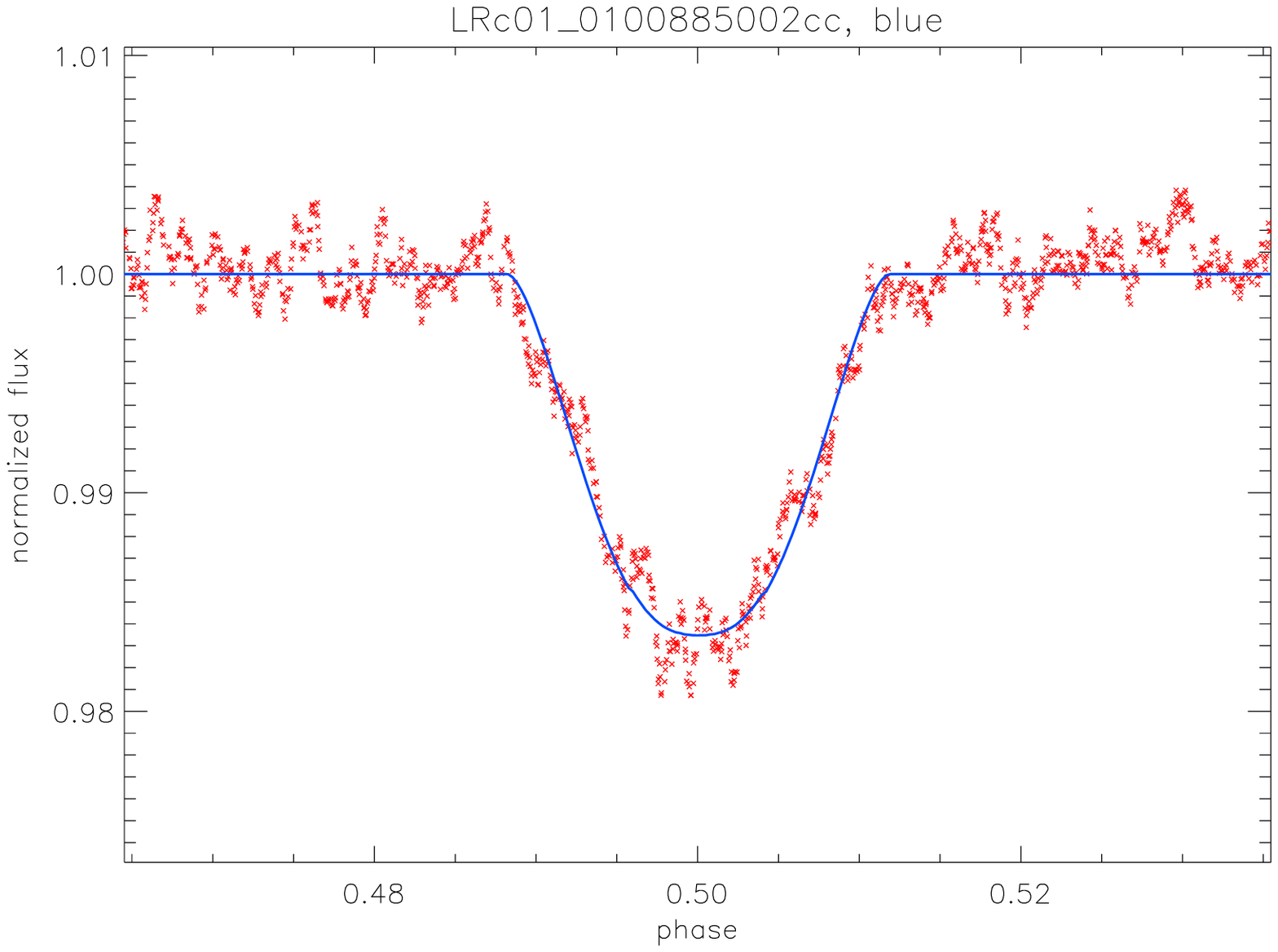}}
\end{minipage}
\caption{Examples of binary systems analysed in the present work. Here and
         in the following figure the light curves were smoothed to reduce
	 noise and clarify visualisation. The index $c$, when added to the
	 CoRoT ID, indicates that the light curve was corrected from
	 short-period oscillations.}
\label{bin_cand}
\end{figure*}

Table~\ref{cand_tab} and Fig.~\ref{bin_cand} show a few examples of our
method applied to light curves of systems classified as eclipsing binaries
or targets having faint background eclipsing binary stars.

\begin{table*}[t!]
\centering
\caption[]{Physical and orbital parameters for CoRoT systems analysed in
           this work. The uncertainties in $a$, $i$, $M_1$, $R_1$, $M_2$,
	   and $R_2$ were estimated as described in Sect.~\ref{errors}.
	   These targets represent confirmed or probable binary systems and
	   they are all commented in Sect.~\ref{res_bin}.}
\label{cand_tab}
{\scriptsize
\begin{tabular}{llcccccccl}
\noalign{\smallskip}\hline\hline\noalign{\smallskip}
CoRoT ID &
Run &
\parbox[c]{0.8cm}{\centering $P$ [days]} &
\parbox[c]{0.65cm}{\centering $a$ [AU]} &
\parbox[c]{0.65cm}{\centering $i$ [deg]} &
\parbox[c]{0.65cm}{\centering $M_1$ [\msun]} &
\parbox[c]{0.65cm}{\centering $R_1$ [\rsun]} &
\parbox[c]{0.85cm}{\centering $M_2$ [\msun]} &
\parbox[c]{0.75cm}{\centering $R_2$ [\rjup]} &
Ref. \\
\noalign{\smallskip}\hline\noalign{\smallskip}
0102787048 & IRa01 & 7.896   & 0.094  $\pm$ 0.008  & 85.1  $\pm$ 0.2  & 1.63 $\pm$ 0.08 & 2.1  $\pm$ 0.2  & 0.16 $\pm$ 0.07 & 2.00 $\pm$ 0.31 & \lbrack 2, 3, 4\rbrack \\
0102811578 & IRa01 & 1.66882 & 0.0339 $\pm$ 0.0008 & 77.1  $\pm$ 0.2  & 1.57 $\pm$ 0.02 & 1.95 $\pm$ 0.04 & 0.32 $\pm$ 0.02 & 3.09 $\pm$ 0.17 & \lbrack 2\rbrack \\
0102815260 & IRa01 & 3.587   & 0.057  $\pm$ 0.006  & $>$~87.0         & 1.71 $\pm$ 0.09 & 2.2  $\pm$ 0.2  & 0.19 $\pm$ 0.08 & 2.14 $\pm$ 0.31 & \lbrack 2, 4\rbrack \\
0102855534 & IRa01 & 21.72   & 0.218  $\pm$ 0.014  & 86.9  $\pm$ 0.2  & 2.48 $\pm$ 0.07 & 4.2  $\pm$ 0.2  & 0.50 $\pm$ 0.05 & 4.49 $\pm$ 0.42 & \lbrack 2, 4\rbrack \\
0100773735 & LRc01 & 4.974   & 0.074  $\pm$ 0.004  & 82.5  $\pm$ 0.2  & 1.98 $\pm$ 0.06 & 2.8  $\pm$ 0.1  & 0.23 $\pm$ 0.04 & 2.42 $\pm$ 0.24 & \lbrack 1\rbrack \\
0100885002 & LRc01 & 11.8054 & 0.136  $\pm$ 0.007  & 85.1  $\pm$ 0.1  & 2.02 $\pm$ 0.02 & 2.9  $\pm$ 0.1  & 0.42 $\pm$ 0.04 & 3.80 $\pm$ 0.31 & \lbrack 1\rbrack \\
0101482707 & LRc01 & 39.89   & 0.270  $\pm$ 0.004  & 88.5  $\pm$ 0.1  & 1.45 $\pm$ 0.01 & 1.73 $\pm$ 0.02 & 0.19 $\pm$ 0.01 & 2.18 $\pm$ 0.09 & \lbrack 1\rbrack \\
0101095286 & LRc01 & 5.053   & 0.088  $\pm$ 0.005  & 70.6  $\pm$ 0.4  & 3.2  $\pm$ 0.1  & 6.7  $\pm$ 0.3  & 0.43 $\pm$ 0.12 & 3.91 $\pm$ 0.92 & \lbrack 1, 3\rbrack \\
0101434308 & LRc01 & 79.95   & 0.41   $\pm$ 0.02   & $>$~89.7         & 1.30 $\pm$ 0.05 & 1.46 $\pm$ 0.08 & 0.18 $\pm$ 0.06 & 2.10 $\pm$ 0.17 & \lbrack 1\rbrack \\
0211660858 & SRc01 & 8.825   & 0.114  $\pm$ 0.008  & 86.3  $\pm$ 0.3  & 2.12 $\pm$ 0.07 & 3.2  $\pm$ 0.2  & 0.46 $\pm$ 0.05 & 4.10 $\pm$ 0.38 & \\
0211654447 & SRc01 & 4.751   & 0.090  $\pm$ 0.005  & 70.9  $\pm$ 0.4  & 3.3  $\pm$ 0.1  & 7.4  $\pm$ 0.3  & 1.04 $\pm$ 0.10 & 9.90 $\pm$ 1.23 & \\
0102755837 & LRa01 & 27.955  & 0.28   $\pm$ 0.02   & 83.8  $\pm$ 0.2  & 3.4  $\pm$ 0.1  & 7.6  $\pm$ 0.4  & 0.48 $\pm$ 0.07 & 4.28 $\pm$ 0.61 & \\
\hline\\[-0.4cm]
\end{tabular}
\begin{flushleft}
\lbrack 1\rbrack\ \citet{Cabreraetal2009};
\lbrack 2\rbrack\ \citet{Carpanoetal2009};
\lbrack 3\rbrack\ \citet{Deegetal2009};
\lbrack 4\rbrack\ \citet{Moutouetal2009}.
\end{flushleft}
}
\end{table*}

According to \citet{SilvaCruz2006}, the present method should consider as
non-planetary candidates only the systems for which the radius of the
secondary companion is larger than 1.5 Jupiter radii. After that
publication, some exoplanets with radius between 1.5 and 2~\rjup\ were
discovered. Thus, in this work, we chose a more conservative value of
$R_2$ = 2~\rjup\ for the lower limit of binary candidates.

The systems that our method classify as binary candidates actually represent
classes of systems with different configurations of the components. There
are several sources of false alarms that can mimic the transit of a planet
in front of the main target. The most common are eclipsing binary systems
with grazing transits, targets having a background eclipsing binary system,
or even triple systems. The present method does not identify the exact
configuration of the system, but provides the information that the observed
transits are likely not caused by a planet. Listed below are our comments
for each case. The window ID of the run (e.g. E1-0288) is also shown as a
complementary identifier.

\subsubsubsection{IRa01 - 0102787048 (E1-0288)}

This system was first classified by \citet{Carpanoetal2009} as a planetary
transit candidate. However, after follow-up observations,
\citet{Moutouetal2009} confirmed that the transit is originated by a
background eclipsing binary and then diluted by the main target. They
estimated a mass-ratio of 0.15 between secondary and primary components of
the binary system. Using the results of our method we have a value of 0.10
$\pm$ 0.04 for the same ratio, which is consistent with a diluted transit of
a more massive object. The fact that no transit is observed in the red
channel also helped to label a non-planetary nature for this target.

\subsubsubsection{IRa01 - 0102811578 (E2-0416)}

System classified by \citet{Carpanoetal2009} as an eclipsing binary. No
information concerning follow-up observations has been published. Our
estimate of mass and radius for the secondary companion confirms its
binary nature.

\subsubsubsection{IRa01 - 0102815260 (E2-2430)}

Also in the list of planetary transit candidates of \citet{Carpanoetal2009},
but afterwards classified by \citet{Moutouetal2009} as a binary system
according to radial-velocity observations. Our estimate for the mass-ratio
is 0.11 $\pm$ 0.04, slightly smaller than the value of 0.17 published by
\citet{Moutouetal2009}, but still consistent with a stellar companion.

\subsubsubsection{IRa01 - 0102855534 (E2-1736)}

Follow-up observations conducted by \citet{Moutouetal2009} indicate that
an eclipsing binary is the main target, changing the planetary nature of the
secondary first suggested by \citet{Carpanoetal2009}. The radial-velocity
measurements indicate a mass-ratio of 0.2, which agrees with our estimate of
0.19 $\pm$ 0.01. Indeed, the mass and radius that we show in
Table~\ref{cand_tab} clearly indicate the stellar nature of the secondary.

\subsubsubsection{LRc01 - 0100773735 (E2-1245)}

Based on radial-velocity observations, \citet{Cabreraetal2009} suggest that
this is a spectroscopic binary or multiple system. Our mass and radius
estimate confirms their conclusion of a non-planetary object causing the
observed transits.

\subsubsubsection{LRc01 - 0100885002 (E2-4653)}

This target is listed in \citet{Cabreraetal2009} as an eclipsing binary,
which is in agreement with the mass and radius estimated using our method.
Transits are observed only in the blue channel, contributing to the
classification of this system as non-planetary.

\subsubsubsection{LRc01 - 0101482707 (E1-2837)}

Also classified by \citet{Cabreraetal2009} as a binary system after
radial-velocity observations and confirmed by our method.

\subsubsubsection{LRc01 - 0101095286 (E1-2376)}

The follow-up of this target with photometric observations unveiled its
binary nature \citep{Cabreraetal2009}, which can be clearly confirmed by our
results of mass and radius for the secondary object.

\subsubsubsection{LRc01 - 0101434308 (E1-3425)}

\citet{Cabreraetal2009}, without doing any follow-up observations, concluded
that this is a binary system given the fact that the transit is
predominantly seen in the blue channel.

In Table~\ref{cand_tab} are also listed our results for the targets
0211660858 (SRc01, E2-0369), 0211654447 (SRc01, E1-1165), and 0102755837
(LRa01, E2-2249). The CoRoT team has not yet published the results of their
analysis for these runs. Nevertheless, they are shown here because
the secondary mass and radius estimated by our
method clearly classify these targets as non-planetary systems.

As one can see in Table~\ref{cand_tab}, for some targets the estimated
radius of the secondary companion is close to the limit of 2~\rjup\ used to
distinguish binaries from possible planetary systems. They were classified
as probable binary systems considering that, at present, all detected
exoplanets for which the radius was derived are as large as 1.8~\rjup\ or
smaller.

Among the nine targets listed in Table~\ref{cand_tab} that were analysed and
published by the CoRoT team, six were classified as binary systems only
after ground-based follow-up, either by radial-velocity measurements, or by
photometric observations, or both. This is an indication that the method
normally used by different CoRoT working groups is perhaps not good enough
to provide a pre-classification of the targets. Our method would exclude the
binary targets by itself, and no time-consuming follow-up observations would
be required for most cases.

\subsection{Characterisation of exoplanetary systems}
\label{res_plan}

\begin{figure*}[p!]
\centering
\begin{minipage}[t]{0.49\textwidth}
\centering
\resizebox{\hsize}{!}{\includegraphics{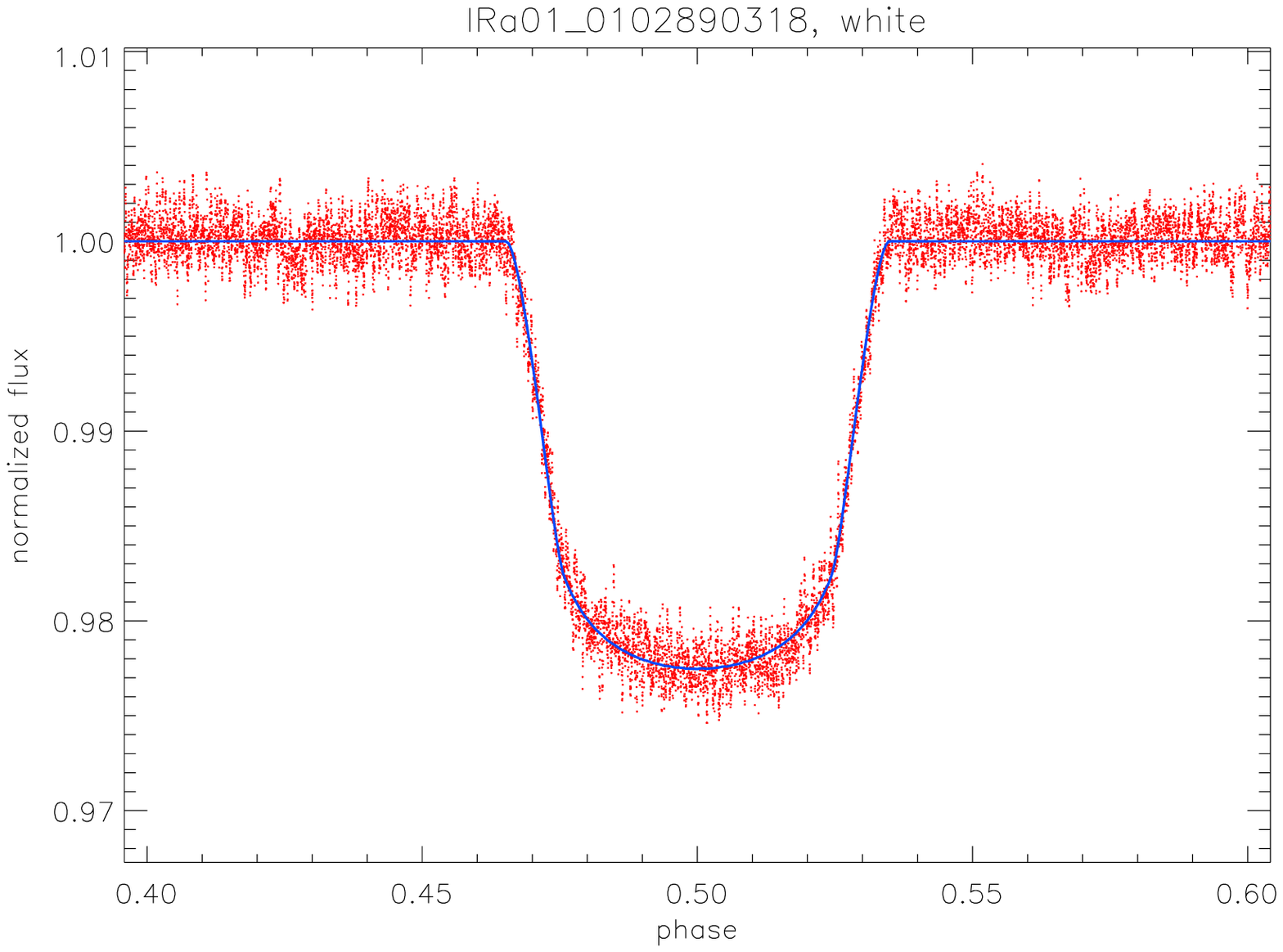}}
\end{minipage}
\begin{minipage}[t]{0.49\textwidth}
\centering
\resizebox{\hsize}{!}{\includegraphics{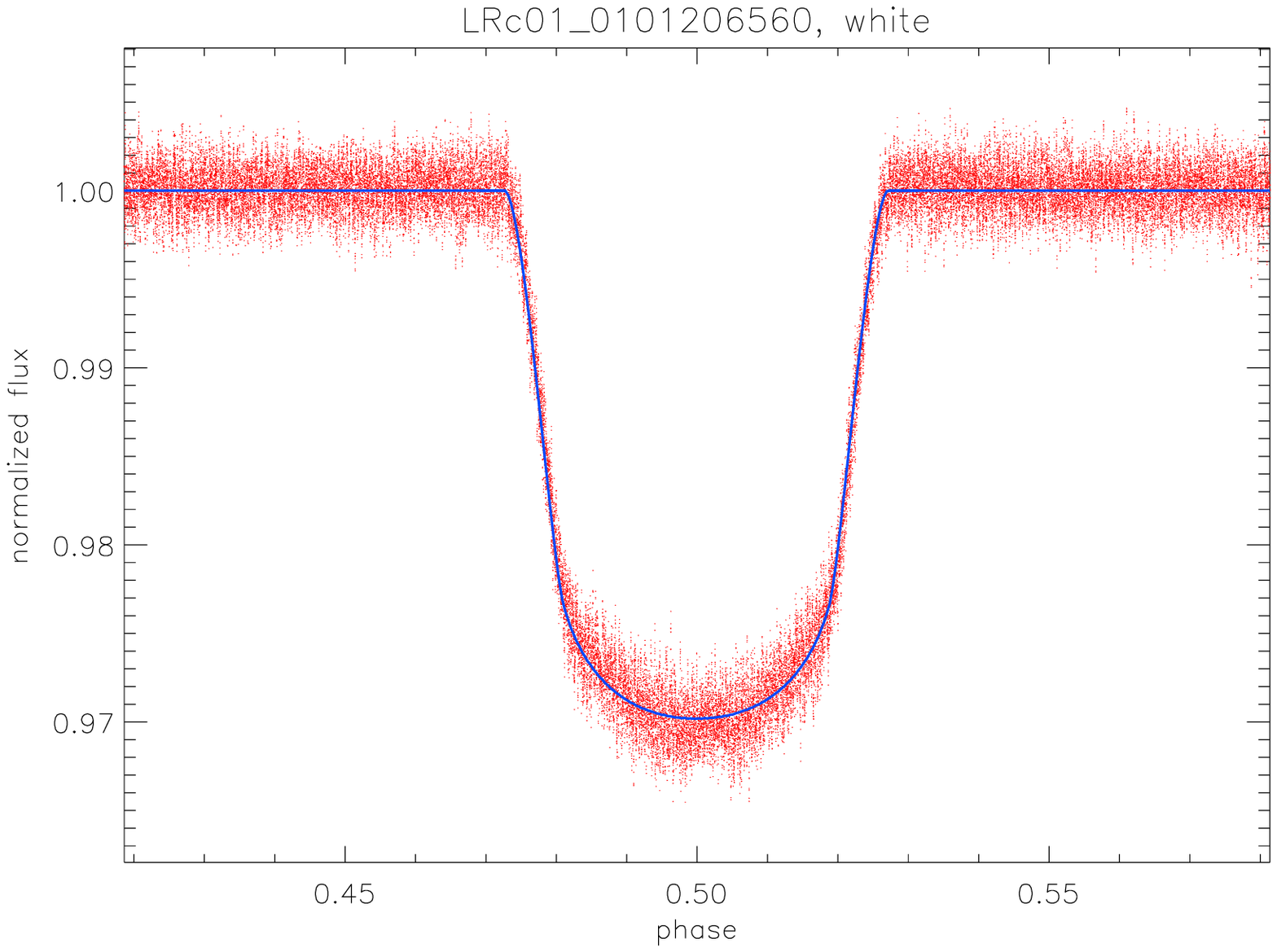}}
\end{minipage} \\[0.80cm]
\begin{minipage}[t]{0.49\textwidth}
\centering
\resizebox{\hsize}{!}{\includegraphics{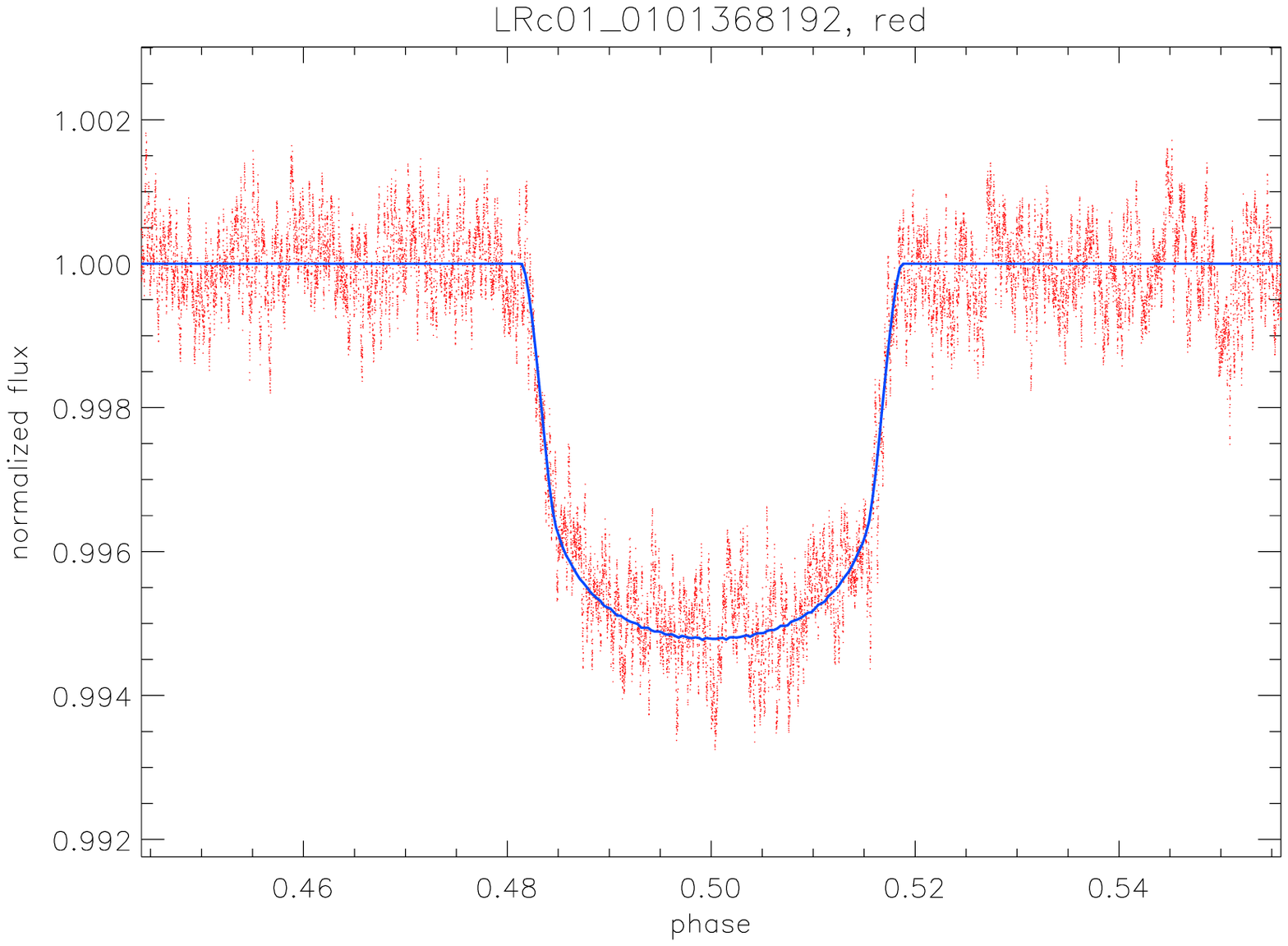}}
\end{minipage}
\begin{minipage}[t]{0.49\textwidth}
\centering
\resizebox{\hsize}{!}{\includegraphics{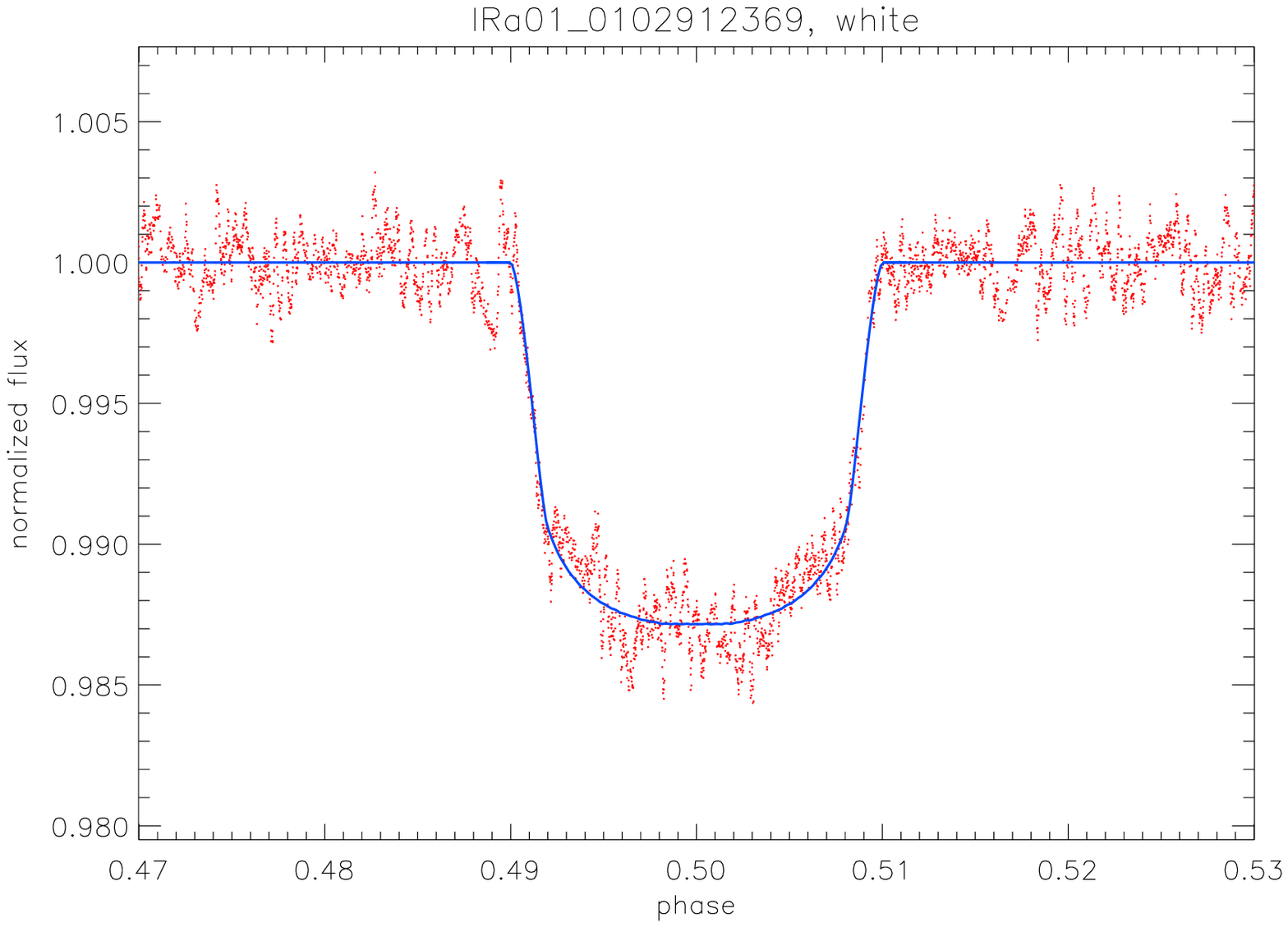}}
\end{minipage} \\[0.80cm]
\begin{minipage}[t]{0.49\textwidth}
\centering
\resizebox{\hsize}{!}{\includegraphics{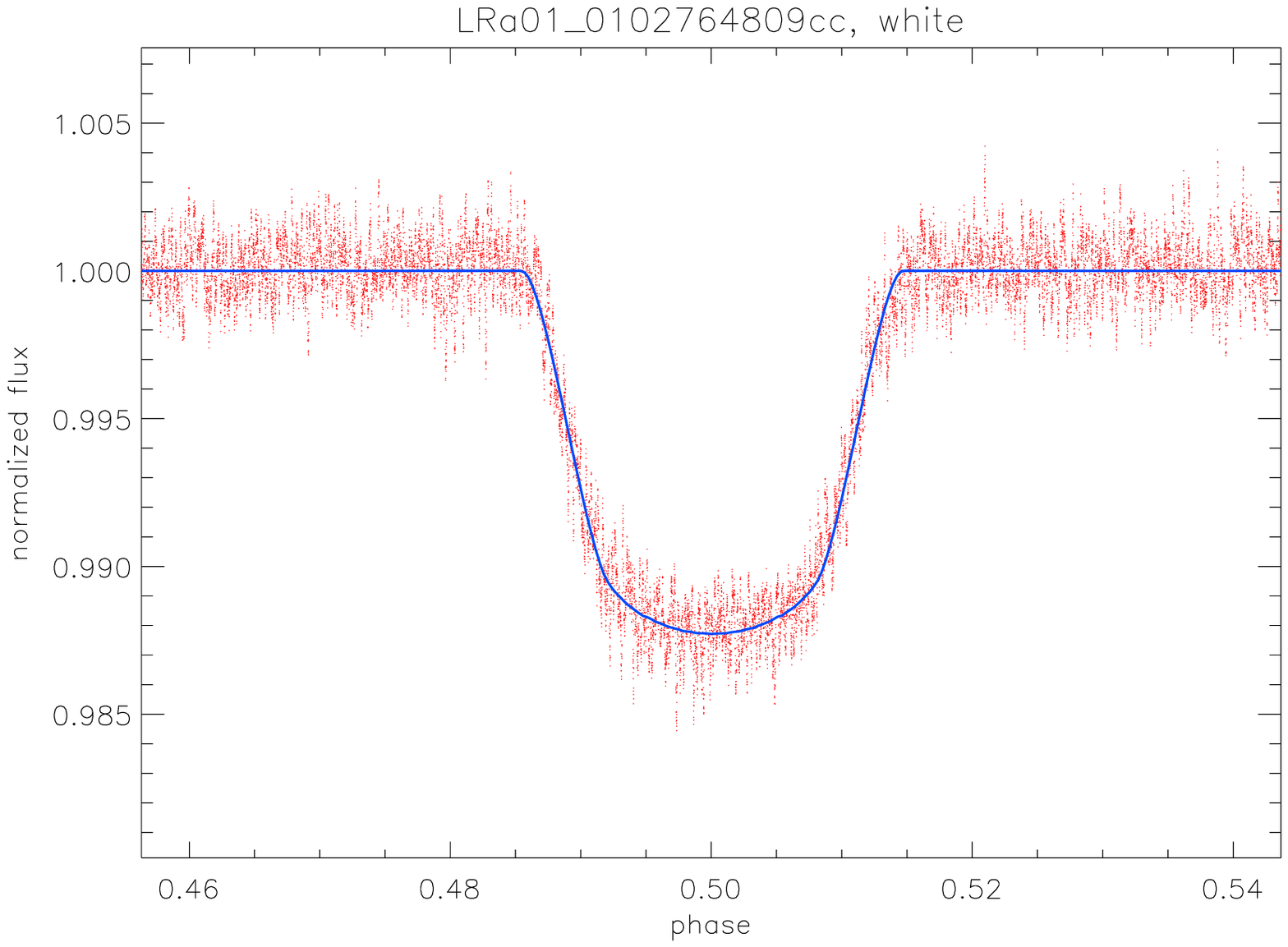}}
\end{minipage}
\begin{minipage}[t]{0.49\textwidth}
\centering
\resizebox{\hsize}{!}{\includegraphics{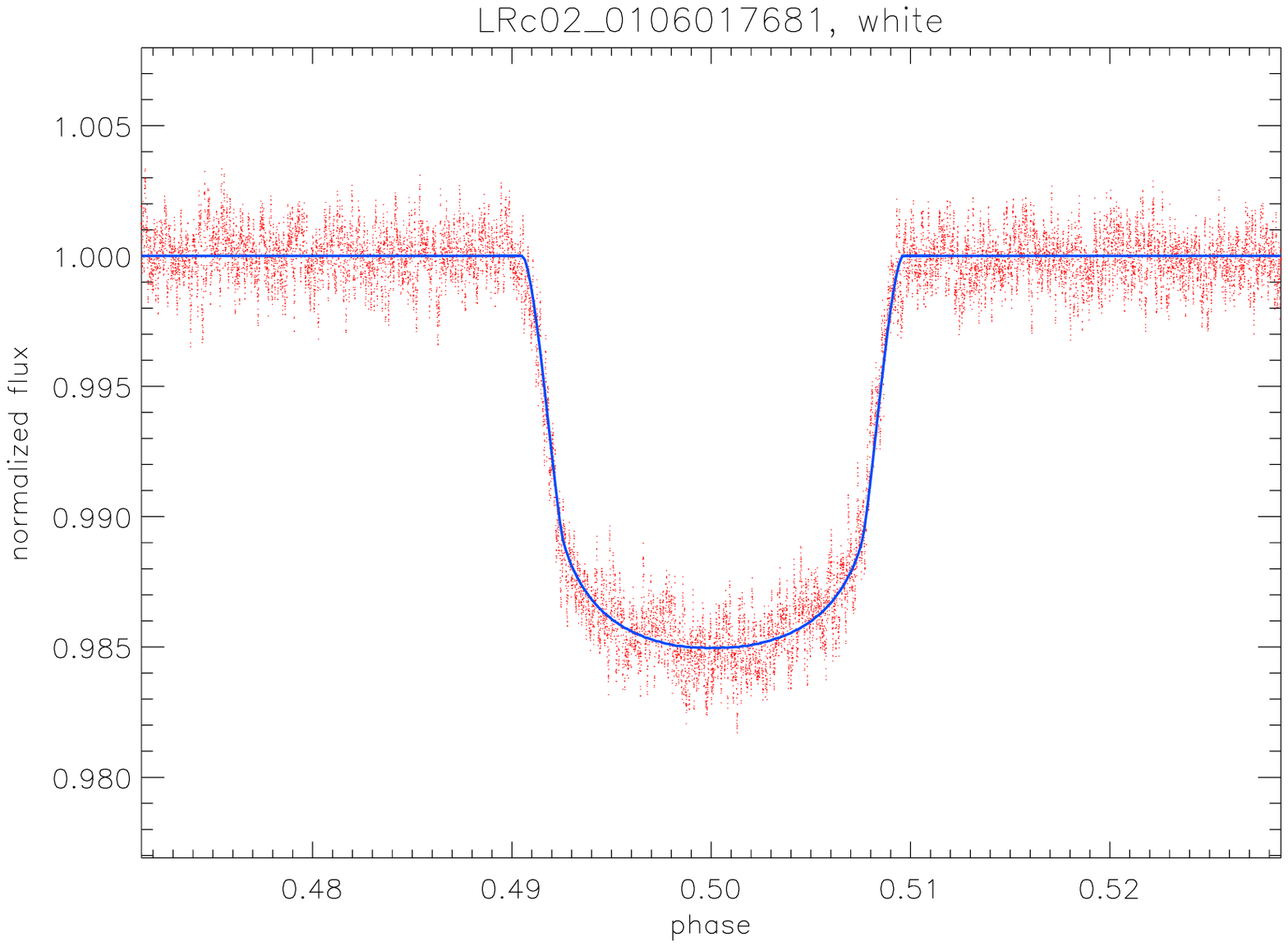}}
\end{minipage}
\caption{The first six CoRoT exoplanetary systems: CoRoT-1 through CoRoT-6,
         respectively, from left to right and from up to bottom.}
\label{corot_fig}
\end{figure*}

Table~\ref{corot_tab} presents our parameter estimates for confirmed CoRoT
exoplanetary systems, from CoRoT-1 through 12, together with the results
presented in the announcement papers. Here the method uses the information
provided by the mass-radius relation fitted to known transiting exoplanets
and to the planets of the solar system (Equation~\ref{eq_mr_2},
Fig.~\ref{massradius}). Figure~\ref{corot_fig} shows the light curves of the
first six CoRoT systems plotted with our best fit.

\begin{table*}[t!]
\centering
\caption[]{Parameters of the first twelve confirmed CoRoT planetary systems
           compared to the results of this work. The period values are from
	   the respective announcement papers. The uncertainties in $a$,
	   $i$, $M_1$, $R_1$, and $R_2$ were estimated as described in
	   Sect.~\ref{errors}.}
\label{corot_tab}
{\scriptsize
\begin{tabular}{lcccccccl}
\noalign{\smallskip}\hline\hline\noalign{\smallskip}
CoRoT ID & Run &
\parbox[c]{0.8cm}{\centering $P$ [days]} &
\parbox[c]{0.65cm}{\centering $a$ [AU]} &
\parbox[c]{0.65cm}{\centering $i$ [deg]} &
\parbox[c]{0.65cm}{\centering $M_1$ [\msun]} &
\parbox[c]{0.65cm}{\centering $R_1$ [\rsun]} &
\parbox[c]{0.85cm}{\centering $R_2$ [\rjup]} &
Ref. \\
\noalign{\smallskip}\hline\noalign{\smallskip}
\parbox[c]{1.4cm}{0102890318 \\ (CoRoT-1)} &
IRa01 & 1.5089557 (64) &
\parbox[c]{2.1cm}{\centering 0.027 $\pm$ 0.002 \\ 0.0254 $\pm$ 0.0004} &
\parbox[c]{1.2cm}{\centering 85.2 $\pm$ 0.5 \\ 85.1 $\pm$ 0.5} &
\parbox[c]{1.3cm}{\centering 1.11 $\pm$ 0.04 \\ 0.95 $\pm$ 0.15} &
\parbox[c]{1.6cm}{\centering 1.18 $\pm$ 0.06 \\ 1.11 $\pm$ 0.05} &
\parbox[c]{1.6cm}{\centering 1.59 $\pm$ 0.13 \\ 1.49 $\pm$ 0.08} &
\parbox[c]{0.6cm}{\lbrack 0\rbrack \\ \lbrack 1\rbrack} \\[-0.15cm]
\noalign{\smallskip}\noalign{\smallskip}\noalign{\smallskip}
\parbox[c]{1.4cm}{0101206560 \\ (CoRoT-2)} &
LRc01 & 1.7429964 (17) &
\parbox[c]{2.1cm}{\centering 0.028 $\pm$ 0.002 \\ 0.0281 $\pm$ 0.0009} &
\parbox[c]{1.2cm}{\centering 87.5 $\pm$ 0.4 \\ 87.8 $\pm$ 0.1} &
\parbox[c]{1.3cm}{\centering 0.90 $\pm$ 0.03 \\ 0.97 $\pm$ 0.06} &
\parbox[c]{1.6cm}{\centering 0.90 $\pm$ 0.04 \\ 0.902 $\pm$ 0.018} &
\parbox[c]{1.6cm}{\centering 1.38 $\pm$ 0.10 \\ 1.465 $\pm$ 0.029} &
\parbox[c]{0.6cm}{\lbrack 0\rbrack \\ \lbrack 2\rbrack} \\[-0.15cm]
\noalign{\smallskip}\noalign{\smallskip}\noalign{\smallskip}
\parbox[c]{1.4cm}{0101368192 \\ (CoRoT-3)} &
LRc01 & 4.256800 (5) &
\parbox[c]{1.8cm}{\centering 0.056 $\pm$ 0.005 \\ 0.057 $\pm$ 0.003} &
\parbox[c]{1.2cm}{\centering 86.3 $\pm$ 0.4 \\ 85.9 $\pm$ 0.8} &
\parbox[c]{1.3cm}{\centering 1.33 $\pm$ 0.07 \\ 1.37 $\pm$ 0.09} &
\parbox[c]{1.6cm}{\centering 1.52 $\pm$ 0.11 \\ 1.56 $\pm$ 0.09} &
\parbox[c]{1.6cm}{\centering 0.98 $\pm$ 0.13 \\ 1.01 $\pm$ 0.07} &
\parbox[c]{0.6cm}{\lbrack 0\rbrack \\ \lbrack 3\rbrack} \\[-0.15cm]
\noalign{\smallskip}\noalign{\smallskip}\noalign{\smallskip}
\parbox[c]{1.4cm}{0102912369 \\ (CoRoT-4)} &
IRa01 & 9.20205 (37) &
\parbox[c]{1.8cm}{\centering 0.087 $\pm$ 0.006 \\ 0.090 $\pm$ 0.001} &
\parbox[c]{1.2cm}{\centering $>$~89.3 \\ $>$~89.915} &
\parbox[c]{1.3cm}{\centering 1.03 $\pm$ 0.04 \\ 1.16 $\pm$ 0.03} &
\parbox[c]{1.6cm}{\centering 1.07 $\pm$ 0.06 \\ 1.17 $\pm$ 0.03} &
\parbox[c]{1.6cm}{\centering 1.07 $\pm$ 0.10 \\ 1.19 $\pm$ 0.06} &
\parbox[c]{0.6cm}{\lbrack 0\rbrack \\ \lbrack 4\rbrack} \\[-0.15cm]
\noalign{\smallskip}\noalign{\smallskip}\noalign{\smallskip}
\parbox[c]{1.4cm}{0102764809 \\ (CoRoT-5)} &
LRa01 & 4.0378962 (19) &
\parbox[c]{2.2cm}{\centering 0.053 $\pm$ 0.003 \\ 0.04947 $\pm$ 0.00029} &
\parbox[c]{1.2cm}{\centering 85.0 $\pm$ 0.2 \\ 86 $\pm$ 1} &
\parbox[c]{1.3cm}{\centering 1.19 $\pm$ 0.04 \\ 1.00 $\pm$ 0.02} &
\parbox[c]{1.6cm}{\centering 1.29 $\pm$ 0.06 \\ 1.19 $\pm$ 0.04} &
\parbox[c]{1.6cm}{\centering 1.37 $\pm$ 0.14 \\ 1.388 $\pm$ 0.047} &
\parbox[c]{0.6cm}{\lbrack 0\rbrack \\ \lbrack 5\rbrack} \\[-0.15cm]
\noalign{\smallskip}\noalign{\smallskip}\noalign{\smallskip}
\parbox[c]{1.4cm}{0106017681 \\ (CoRoT-6)} &
LRc02 & 8.886593 (4) &
\parbox[c]{2.2cm}{\centering 0.083 $\pm$ 0.006 \\ 0.0855 $\pm$ 0.0015} &
\parbox[c]{1.2cm}{\centering 89.4 $\pm$ 0.4 \\ 89.1 $\pm$ 0.3} &
\parbox[c]{1.3cm}{\centering 0.96 $\pm$ 0.05 \\ 1.05 $\pm$ 0.05} &
\parbox[c]{1.6cm}{\centering 0.98 $\pm$ 0.06 \\ 1.025 $\pm$ 0.026} &
\parbox[c]{1.6cm}{\centering 1.06 $\pm$ 0.10 \\ 1.166 $\pm$ 0.035} &
\parbox[c]{0.6cm}{\lbrack 0\rbrack \\ \lbrack 6\rbrack} \\[-0.15cm]
\noalign{\smallskip}\noalign{\smallskip}\noalign{\smallskip}
\parbox[c]{1.4cm}{0102708694 \\ (CoRoT-7)} &
LRa01 & 0.853585 (24) &
\parbox[c]{2.2cm}{\centering 0.018 $\pm$ 0.005 \\ 0.01720 $\pm$ 0.00029} &
\parbox[c]{1.2cm}{\centering 78.2 $\pm$ 1.5 \\ 80.1 $\pm$ 0.3} &
\parbox[c]{1.3cm}{\centering 0.98 $\pm$ 0.17 \\ 0.93 $\pm$ 0.03} &
\parbox[c]{1.6cm}{\centering 1.01 $\pm$ 0.23 \\ 0.87 $\pm$ 0.04} &
\parbox[c]{1.6cm}{\centering 0.15 $\pm$ 0.07 \\ 0.150 $\pm$ 0.008} &
\parbox[c]{0.6cm}{\lbrack 0\rbrack \\ \lbrack 7\rbrack} \\[-0.15cm]
\noalign{\smallskip}\noalign{\smallskip}\noalign{\smallskip}
\parbox[c]{1.4cm}{0101086161 \\ (CoRoT-8)} &
LRc01 & 6.21229 (3) &
\parbox[c]{2.2cm}{\centering 0.067 $\pm$ 0.004 \\ 0.063 $\pm$ 0.001} &
\parbox[c]{1.2cm}{\centering 86.7 $\pm$ 0.1 \\ 88.4 $\pm$ 0.1} &
\parbox[c]{1.3cm}{\centering 1.06 $\pm$ 0.03 \\ 0.88 $\pm$ 0.04} &
\parbox[c]{1.6cm}{\centering 1.11 $\pm$ 0.05 \\ 0.77 $\pm$ 0.02} &
\parbox[c]{1.6cm}{\centering 0.88 $\pm$ 0.09 \\ 0.57 $\pm$ 0.02} &
\parbox[c]{0.6cm}{\lbrack 0\rbrack \\ \lbrack 8\rbrack} \\[-0.15cm]
\noalign{\smallskip}\noalign{\smallskip}\noalign{\smallskip}
\parbox[c]{1.4cm}{0105891283 \\ (CoRoT-9)} &
LRc02 & 95.2738 (14) &
\parbox[c]{2.2cm}{\centering 0.39 $\pm$ 0.02 \\ 0.407 $\pm$ 0.005} &
\parbox[c]{1.2cm}{\centering $>$~89.87 \\ $>$~89.95 } &
\parbox[c]{1.3cm}{\centering 0.85 $\pm$ 0.03 \\ 0.99 $\pm$ 0.04} &
\parbox[c]{1.6cm}{\centering 0.84 $\pm$ 0.04 \\ 0.94 $\pm$ 0.04} &
\parbox[c]{1.6cm}{\centering 0.93 $\pm$ 0.08 \\ 1.05 $\pm$ 0.04} &
\parbox[c]{0.6cm}{\lbrack 0\rbrack \\ \lbrack 9\rbrack} \\[-0.15cm]
\noalign{\smallskip}\noalign{\smallskip}\noalign{\smallskip}
\parbox[c]{1.4cm}{0100725706 \\ (CoRoT-10)} &
LRc01 & 13.2406 (2) &
\parbox[c]{2.2cm}{\centering 0.108 $\pm$ 0.005 \\ 0.1055 $\pm$ 0.0021} &
\parbox[c]{1.2cm}{\centering 88.0 $\pm$ 0.1 \\ 88.6 $\pm$ 0.2} &
\parbox[c]{1.3cm}{\centering 0.94 $\pm$ 0.03 \\ 0.89 $\pm$ 0.05} &
\parbox[c]{1.6cm}{\centering 0.95 $\pm$ 0.04 \\ 0.79 $\pm$ 0.05} &
\parbox[c]{1.6cm}{\centering 1.15 $\pm$ 0.10 \\ 0.97 $\pm$ 0.07} &
\parbox[c]{0.6cm}{\lbrack 0\rbrack \\ \lbrack 10\rbrack} \\[-0.15cm]
\noalign{\smallskip}\noalign{\smallskip}\noalign{\smallskip}
\parbox[c]{1.4cm}{0105833549 \\ (CoRoT-11)} &
LRc02 & 2.994330 (11) &
\parbox[c]{2.2cm}{\centering 0.044 $\pm$ 0.002 \\ 0.044 $\pm$ 0.005} &
\parbox[c]{1.5cm}{\centering 83.2 $\pm$ 0.2 \\ 83.17 $\pm$ 0.15} &
\parbox[c]{1.3cm}{\centering 1.24 $\pm$ 0.04 \\ 1.27 $\pm$ 0.05} &
\parbox[c]{1.6cm}{\centering 1.37 $\pm$ 0.06 \\ 1.37 $\pm$ 0.03} &
\parbox[c]{1.6cm}{\centering 1.32 $\pm$ 0.13 \\ 1.43 $\pm$ 0.03} &
\parbox[c]{0.6cm}{\lbrack 0\rbrack \\ \lbrack 11\rbrack} \\[-0.15cm]
\noalign{\smallskip}\noalign{\smallskip}\noalign{\smallskip}
\parbox[c]{1.4cm}{0102671819 \\ (CoRoT-12)} &
LRa01 & 2.828042 (13) &
\parbox[c]{2.2cm}{\centering 0.040 $\pm$ 0.003 \\ 0.0402 $\pm$ 0.0009} &
\parbox[c]{1.2cm}{\centering 85.7 $\pm$ 0.2 \\ 85.5 $\pm$ 0.8} &
\parbox[c]{1.3cm}{\centering 1.02 $\pm$ 0.05 \\ 1.08 $\pm$ 0.08} &
\parbox[c]{1.6cm}{\centering 1.06 $\pm$ 0.07 \\ 1.12 $\pm$ 0.10} &
\parbox[c]{1.6cm}{\centering 1.33 $\pm$ 0.15 \\ 1.44 $\pm$ 0.13} &
\parbox[c]{0.6cm}{\lbrack 0\rbrack \\ \lbrack 12\rbrack} \\
\noalign{\smallskip}\hline\\[-0.4cm]
\end{tabular}
\begin{flushleft}
\lbrack 0\rbrack\ this work;
\lbrack 1\rbrack\ \citet{Bargeetal2008};
\lbrack 2\rbrack\ \citet{Alonsoetal2008};
\lbrack 3\rbrack\ \citet{Deleuiletal2008};
\lbrack 4\rbrack\ \citet{Aigrainetal2008};
\lbrack 5\rbrack\ \citet{Raueretal2009}; \\
\lbrack 6\rbrack\ \citet{Fridlundetal2010};
\lbrack 7\rbrack\ \citet{Legeretal2009};
\lbrack 8\rbrack\ \citet{Bordeetal2010};
\lbrack 9\rbrack\ \citet{Deegetal2010};
\lbrack 10\rbrack\ \citet{Bonomoetal2010}; \\
\lbrack 11\rbrack\ \citet{Gandolfietal2010};
\lbrack 12\rbrack\ \citet{Gillonetal2010}
\end{flushleft}
}
\end{table*}

The results of a detailed comparison between our estimates and those of the
announcement papers are shown in Table~\ref{regr_tab}. This table lists the
coefficients of weighted linear regressions obtained for $a$, $i$, $M_1$,
$R_2$, and $R_2$, where the weights $1/\sigma^2$ were used for $\sigma$
representing the errors that we estimated for these parameters. At first, we
computed the coefficients using the first twelve CoRoT systems and no
systematic difference was found within 2$\sigma$. Within 1$\sigma$, however,
the agreement is not so good, specially for $M_1$ and $R_1$. This is most
due to discrepancies in the comparison of CoRoT-8, as one can observe in
Table~\ref{corot_tab}. The data points in the light curve of this system has
a dispersion of 0.017, which is much larger than those of the other CoRoT
systems: $\sim$0.007 or smaller. Indeed, if the parameters of CoRoT-8 are
not included in the regressions, the agreement between our results and those
of the announcement papers is much better, standing within 1$\sigma$ mostly.
An extra caution when dealing with noisy light curves is, therefore,
recommended. Light curves with small-depth or long-period transits (such as
CoRoT-7 and 9, respectively) also produce larger uncertainties in the
parameters and should be carefully analysed.

\begin{table}[t!]
\centering
\caption[]{Coefficients of the linear regression obtained in the comparison
           of the parameters listed in Table~\ref{corot_tab}, with CoRoT-8
	   included or not (see Sect.~4.2). The angular and linear
	   coefficients ($\alpha$, $\beta$), the dispersion around the fit
	   ($rms$), and the correlation coefficient ($cc$) are shown for
	   both scenarios.}
\label{regr_tab}
{\footnotesize
\begin{tabular}{lcccc}
\noalign{\smallskip}\hline\hline\noalign{\smallskip}
      & $\alpha$	& $\beta$	      & $rms$  & $cc$ \\
\noalign{\smallskip}\hline\noalign{\smallskip}
\multicolumn{5}{l}{For the first twelve CoRoT systems:} \\[0.05cm]
$a$   & 0.98 $\pm$ 0.02 & 0.0016 $\pm$ 0.0010 & 0.0017 & 1.00 \\
$i$   & 0.91 $\pm$ 0.12 & \, \,7 $\pm$ 10     & 0.68   & 0.97 \\
$M_1$ & 0.55 $\pm$ 0.27 &   0.47 $\pm$ 0.28   & 0.11   & 0.68 \\
$R_1$ & 0.66 $\pm$ 0.19 &   0.38 $\pm$ 0.19   & 0.12   & 0.84 \\
$R_2$ & 0.88 $\pm$ 0.09 &   0.11 $\pm$ 0.10   & 0.13   & 0.94 \\
\hline\noalign{\smallskip}
\multicolumn{5}{l}{CoRoT-8 not included:} \\[0.05cm]
$a$   & 0.97 $\pm$ 0.02 & 0.0017 $\pm$ 0.0009 & 0.0015 & 1.00 \\
$i$   & 0.97 $\pm$ 0.06 &      2 $\pm$ 5      & 0.39   & 0.98 \\
$M_1$ & 0.72 $\pm$ 0.28 &   0.26 $\pm$ 0.29   & 0.10   & 0.73 \\
$R_1$ & 0.84 $\pm$ 0.16 &   0.17 $\pm$ 0.16   & 0.09   & 0.91 \\
$R_2$ & 0.95 $\pm$ 0.07 &   0.02 $\pm$ 0.07   & 0.09   & 0.97 \\
\hline
\end{tabular}
}
\end{table}

In Table~\ref{corot_tab}, all planets detected in the CoRoT field have
$R_2 < 1.5$~\rjup, also including our estimate for CoRoT-1 if we take its
uncertainty into account. This result supports our suggestion that the
present method can be used to characterise not only systems for which the
radius of the secondary companion is larger than 2~\rjup, but also those
having $R_2$ smaller than 1.5~\rjup, helping as a complementary approach in
the search of promising candidates for radial-velocity follow-up. The
intermediate situations, when $R_2$ stands between 1.5 and 2~\rjup, are not
conclusive.

%
%
\section{Conclusions}
\label{concl}

We have presented a method that provides a good estimate of some physical
and orbital parameters of a transiting system, such as the mass and radius
of the secondary companion. Applied to transiting light curves, the method
will exclude cases most probably related to low-mass stars in a binary
system, instead of a planet. In other words, our method is able to exclude
systems that at first may be considered as good planetary candidates but
that afterwards would have their binary nature unveiled, without making use
of time-consuming ground-based measurements normally conducted to complement
the observations.

We note that the method does not, by itself, determine the real nature of
the secondary object (whether it is a binary companion or not). Instead, it
identifies and characterises {\it probable candidates} for binary systems,
which will help to reduce the huge number of targets initially available
and to create a list of priority stars, still candidates for planetary
systems, to be monitored with radial-velocity measurements. We do not
discard other methods, however, which can be used to complement our
approach.

The method was also applied to twelve CoRoT targets confirmed as planetary
systems, showing that the estimated radii of the secondary companions (as
well as other orbital parameters) are in very good agreement with the
results published by the respective announcement papers. This means that our
model can also be used in the characterisation of possible exoplanetary
systems, specially when $R_2$ is smaller than or of the order of 1.5~\rjup.
No conclusions could be drawn concerning the radius range $1.5 < R_2
< 2$~\rjup.

Our model is useful not only to be applied to CoRoT light curves that have
been or will be released by the mission, but also to data of other present
or future missions based on photometric observations of transiting systems
that involve a large sample of targets (such as the Kepler mission).

%
%
\acknowledgements
We thank the financial support from Funda\c{c}\~ao de Amparo \`a Pesquisa do
Estado de S\~ao Paulo (FAPESP) in the form of a grant (2006/50654-3) and a
fellowship (2008/03855-9). We also thanks the Instituto Nacional de
Pesquisas Espaciais (INPE) for its support.


\bibliographystyle{aastex}
\bibliography{daSilvaValio2011}

\end{document}